\documentstyle[12pt]{article}
\newcommand{\gsim}{\raisebox{-.5ex}{\mbox{$\, \stackrel{>}{\sim}$}}}

\begin{document}
\begin{center}{

 P-and-T--VIOLATION TESTS WITH POLARIZED RESONANCE NEUTRONS

           V.~E.~Bunakov, Y.~Novikov\\
Petersburg Nuclear Physics Institute, 188350 Gatchina, Russia\\
             }
\end{center}

e-mail: inovikov@snoopy.phys.spbu.ru

\vspace{2cm}



\noindent {\bf Abstract}:
The enhancements of CP-violating effects in resonance
neutron transmission through polarized targets are studied for 2 possible
versions of experiment. The importance is stressed of error analysis and of
pseudomagnetic effects' compensation.

\newpage

\noindent {\bf 1. Introduction.}  It was shown [1-3] about 15 years ago that
CP-violation effects in
transmission of polarized neutrons through the polarized target might be
enhanced in the vicinity of p-resonances by 5-6 orders of magnitude.
Originally it was suggested to measure the difference in  transmission of
neutrons with spins parallel ($N_+$) and antiparallel ($N_+$) to the vector
$\vec{k}_n\times \vec{I}$
($\vec{k}_n$ and $\vec{I}$ are the neutron momentum and the target spin):
\begin{equation}
\eta_T=\frac{N_+-N_-}{N_++N_-}\approx 2\frac{\sigma_+-\sigma_-}
{\sigma_++\sigma_-}
\end{equation}
Here $N_+$ and $N_-$ are the numbers of neutrons with the corresponding
helicities transmitted through the polarized target sample, $\sigma_+$ and
$\sigma_-$ are the corresponding total neutron cross-sections.
However, it was pointed [4] that without the special precautions the nuclear
pseudo-magnetic precession of neutron spin together with the precession
induced by the P-violating interactions would  give rise to numerous effects
camouflaging the CP-violating ones. As a possible remedy of this nuisance
it was suggested [4] to compensate the nuclear pseudo-magnetic field by the
external magnetic field in order to nullify the neutron spin rotation angle
$\phi$. However, in order to measure the CP-violating interaction with
the reasonable accuracy (about $10^{-4}$ of the P-violating one) it was
necessary to check the spin rotation angle with the precision of about
$ 10^{-7}$ rad [4].

In order to circumvent the above difficulties Stodolsky [5] suggested to
measure the difference $N_{+-}-N_{-+}$, where $N$ is the number of neutrons
transmitted through the target and the subscript indices mean the neutron
helicity before and after the transmission. Consider the polarized
neutron scattering amplitude of the form:
\begin{equation}
f = A + p_tB\cdot(\vec{s}_n \cdot \vec{I}) +
C\cdot(\vec{s}_n \cdot \vec{k}_n) + p_tD\cdot \vec{s}_n \cdot [\vec{k}_n
\times \vec{I}]
\end{equation}
where $\vec{s}_n$ is neutron spin, $p_t$ is the target degree of polarization,
$A$ and $B$ are the spin-independent and spin-dependent parts of the
strong interaction amplitude, $D$ is the P- and CP-violating
interaction amplitude, respectively. The term $C$ contains contributions
from both weak P-violating and strong interaction (from the term of the
type $(\vec{s}_n\cdot \vec{k}_n)(\vec{k}_n \cdot \vec{I})$ in scattering
amplitude -see e.g. [6]).

Stodolsky demonstrated that the difference 
\begin{equation}
N_{+-}-N_{-+} \sim Im(DB^*)
\end{equation}
is free from the above camouflaging effects. It is well-known that in order
to improve the accuracy it is preferable to measure the relative values, i.e.
to normalize the above difference. Although Stodolsky never bothered to
introduce this normalization, it seems natural to consider the ratio:
\begin{equation}
T = \frac{N_{+-} - N_{-+}}{N_{+-} + N_{-+}}
\end{equation}

A few years later Serebrov [7] suggested to measure the quantity:
\begin{equation}
X = \frac{(N_{++}-N_{--}) + (N_{+-} - N_{-+})}
{(N_{++}-N_{--}) - (N_{+-} - N_{-+})}
\end{equation}
One can easily see that 
\begin{equation}
X = 1 + 2 \frac{N_{+-} - N_{-+}}{(N_{++}-N_{--}) - (N_{+-} - N_{-+})}
\end{equation}
The actual CP-violating effect causes the deviation of $X$ from unity.
Therefore the actually measured quantity $\tilde{X}$ 
\begin{equation}
\tilde{X} = \frac{N_{+-} - N_{-+}}{(N_{++}-N_{--}) - (N_{+-} - N_{-+})}
\end{equation}
is simply the one suggested by Stodolsky, but normalized in a rather odd
manner.

The main point is that up to now nobody cared to do the analysis of the
energy dependence of the quantities $T$ or $\tilde{X}$ in the manner it
was done for the originally considered CP-violating quantity $\eta_T$ in
refs. [1-3]. Indeed , all the
quantities in the numerators and denominators of $T$ and $\tilde{X}$ contain
various combinations of real and imaginary parts of all the four amplitudes
($A$, $B$, $C$ and $D$) in Eq. (2). Most of them
show a rather complicated energy dependence (see e.g. [3,6,8]) in the
resonance region. Some of them not only vary in magnitude, but even change
their sign. This means
that up to now one does not know whether the suggested values $T$ and
$\tilde{X}$ are really enhanced and what is the magnitude of this enhancement,
if any. Investigation of these problems is the main point of our present
publication. For the time being we are not going to consider the false
effects arising from the difference of the polarizing and the analyzing power
of polarizer and analyzer. We shall also restrict ourselves with cases of
"ideal geometry" when the incident beam polarization is either parallel
or anti-parallel to the neutron momentum.\\

\noindent {\bf 2. Analysis of $T$.} In order to obtain the expressions
for the relative
quantities of interest in terms of the energy-dependent complex amplitudes
$A$, $B$, $C$ and $D$ of Eq. (2), one might use the method developed in
ref. [9]. Introducing the spin density matrix and the evolution operators of
ref. [9], one obtains the expression for $T$:
\begin{equation}
T =  2 \frac{Im(DB^*)}{|D|^2 + |B|^2}
\end{equation}
The expressions for complex amplitudes $D(E)$ and $B(E)$ are obtained using
the methods of ref. [3] (see also [6], [8]).
The main contribution to the T-noninvariant amplitude $D$ in the vicinity
of the $p_{I+1/2}$ -resonance comes from the term coupling this resonance
with the corresponding $s_{I+1/2}$-resonance:
\begin{equation}
D \approx \frac{\gamma^{s}_{I+1/2}}{(E-E^s_{I+1/2}) + \imath
\Gamma_s/2} V_T \frac{\gamma^{p}_{I+1/2}}{(E-E^p_{I+1/2}) +
\imath \Gamma_p/2}.
\end{equation}
In the optimal cases (like $La$ target) these $s_{I+1/2}$ and $ p_{I+1/2}$
resonances contribute equally to the strong amplitude $B$ in this energy
region. Taking into account all the other resonances would only lead
to some numerical changes, while the general qualitative picture would be
the same. Therefore we consider:
\begin{equation}
B \approx \frac{\gamma^s_{I + 1/2} \gamma^s_{I + 1/2}}
{(E-E^s_{I + 1/2}) + \imath \Gamma_s/2} + \frac{\gamma^p_{I + 1/2}
\gamma^p_{I + 1/2}}{(E-E^p_{I + 1/2}) + \imath \Gamma_p/2},\nonumber \\
\end{equation}

Inserting these expressions into
eq. (8) we see that the quantity $T(E)$  in the vicinity of the
p-wave resonance energy $E_p$ is:
\begin{equation}
T(E)\approx -2\frac{\gamma^p_{I+1/2}}{\gamma^s_{I+1/2}} \cdot
\frac{V_T \cdot \Gamma_p}{(E-E_p)^2+\Gamma_p ^2/4}
\end{equation}
Here $\Gamma_p$ stands for the p-resonance
total width, while $V_T$ is the matrix
element of CP-violating interaction causing the transition between the p-
and s-resonance states. Further on in our numerical calculations we shall
assume the ratio of the CP-violating interaction strength to the P-violating
one to be $10^{-4}$ (i.e. $V_T/V_P=10^{-4}$). The quantities $\gamma^{s,p}_J$
stand for the
neutron width amplitudes of the s- and p-resonances with spin $J=I+1/2$.
The sign of the effect is defined by the signs of $\gamma$'s and $V_T$.
For the sake of simplicity we shall choose them in our numerical calculations
so that the net effect is positive.

We observe in Eq. (11) the resonance enhancement of the effect typical for
all the
symmetry-breaking effects in nuclear reactions (see [3,8]). In order to
see explicitly the "dynamical enhancement", which is also typical for these
effects, one might cast the value of $T$ in this maximum in the following
form:
\begin{equation}
T \left( E = E^p_{I+1/2} \right) \approx
\frac{\gamma^p_{I+1/2}}{\gamma^s_{I+1/2}} \cdot \frac{V_T}{d}
\cdot \frac{d}{\Gamma}
\end{equation}
Here $d$ and $\Gamma$ stands for the average resonance spacing and
total width.
It is instructive to remind that the corresponding expression for the
maximal value of the quantity $\eta_T$ obtained in [1-3] was:
\begin{equation}
\eta_T(E=E_p)\approx\frac{\gamma^s}{\gamma^p} \cdot \frac{V_T}{d}
\cdot (\frac{d}{\Gamma})^2
\end{equation}
Comparing Eqs.(12) and (13), one can see in both cases the presence of the
dynamical
enhancement factors $V_T/d\approx F_T\cdot 10^3$ ($F_T$ is the strength of
the CP-violating interaction relative to the strong interaction one) and
of the resonance enhancement factors $d/\Gamma\approx 10^3$ coming from the
fact that the effect is proportional to the time $\tau\sim(1/\Gamma)$ spent
by the incident neutron in the CP-violating field of the target. We also see
the presence of the "entrance channel hindrance" factor (see [3,8])
$\gamma^p/\gamma^s\approx 10^{-3}$ typical for all the low energy scattering
experiments with P-violation. However, the resonance enhancement factor
enters the quantity $\eta_T$ quadratically, while $T$ contains it only
linearly. Therefore the net enhancement of the $T$ quantity is only by
a factor of $10^3$ instead of the $10^5\div 10^6$ factor in $\eta_T$.

These conclusions are illustrated in Fig. 1a, where the energy behavior
of the quantity $T(E)$ is shown for the particular case of the famous
$La$ p-resonance at $E_p=0.75$ eV.

Consider now a very important problem of the optimal choice of the target
sample thickness. One should mind that in the case of $\eta_T$ value,
likewise in the case of the longitudinal polarization $P$ caused by the
P-violating weak interaction, the correct expression for the experimentally
measured ratio can be written as follows (see e.g. [10], [3] and [11]):
\begin{equation}
\frac{N_+-N_-}{N_++N_-}\approx\frac{\sigma_+-\sigma_-}{2}\cdot x\cdot\rho
\end{equation}
where $x$ is the target sample thickness and $\rho$ is the density of nuclei
in this sample. Since the experimentally observed effect is linear in target
thickness, it seems that one should choose the thickest target possible.
However, the neutron countings $N_{\pm}$ decrease exponentially with $x$.
Therefore the statistical relative error of measuring each $N$ value 
\begin{equation}
\frac{\delta N}{N}=\frac{1}{\sqrt{N}}=\frac{1}{\sqrt{N_0}}e^{x\rho\sigma/2}
\end{equation}
also increases exponentially with $x$ ($N_0$ here stands for the number of
polarized neutrons incident on the target). In order to find the optimal
target thickness $x_0$ one should estimate the relative error
$\sigma_{\eta_T} /\eta_T$ of the quantity in the l.h.s. of Eq. (14) and define
its
minimum (by equating the x-derivative of the relative error to zero). In this
way one obtains that the optimal thickness in the case of $\eta_T$ quantity
is $2\lambda=2/\sigma\rho$ (here $\lambda$ stands for the mean free path of
the neutron in the target sample). It is only by choosing the optimal
$x_0$ that one obtains the last line in Eq. (1).

The relative error of the quantity $T$ looks more complicated. One
can easily see that the main contribution to it comes from the relative error
of the numerator in $T$:
\begin{equation}
\frac{\sigma_T}{T} \approx \frac{e^{\frac{Im(A)}{Im(f)}
\frac{x}{\lambda}}}{\sqrt{2N_0}} \cdot
\frac{\left| q \right|}{\sin(\theta)} \cdot
\frac{1}{\sqrt{ ch^2
\left( \frac{Im(q)}{Im(f)} \frac{x}{\lambda} \right) - \cos^2
\left( \frac{Re(q)}{Im(f)} \frac{x}{\lambda} \right) }} \cdot
\sqrt{\frac{|D|^2+|B|^2}{Im^2(DB^*)}}
\end{equation}
Here $\lambda$ is the neutron mean free path and the (complex) quantity
$q$ is defined as:
\begin{equation}
q = \sqrt{ \frac14 \sin^2(\theta) B^2 + \frac14 \sin^2(\theta) D^2 +
\frac14 (C + \cos(\theta) B)^2 }
\end{equation}
The angle between the target polarization and neutron momentum vectors is
denoted as $\theta$. The $\sin(\theta)$ behaviour of Eqs. (16) reflects
the fact that the CP-violating term in the amplitude (2) is proportional to
$\sin(\theta)$. Therefore irrespective of the value of $D$ the CP-violating
effects disappear for $\theta\approx 0$ and the relative error goes to
infinity. The dependence of Eq. (16) on the target thickness $x$ is
complicated by the periodic $\cos^2$ oscillations. The physical origin
of those oscillations is the pseudo-magnetic neutron spin rotation, discussed
in ref. [4] - the
neutron spin performs about a hundred rotations per mean free path in the
target sample. The explicit dependence of the relative error (16) on the
target thickness is shown in Fig. 1b for the case of the same p-resonance
in $La$. The total number of the polarized neutrons $N_0$ incident on the
target was somewhat arbitrary chosen to be $10^{18}$.

One can see from
Fig. 1b that the first minimum of the relative error is located about
$x\approx 10^{-2}\lambda$. However, a slight change of $x$ increases the
relative error by orders of magnitude, which makes the analysis
of the experimental results practically impossible.

This forces us to return to our initial idea [4] of compensating the
pseudomagnetic precession by the external magnetic field. This field
can be formally taken into account by substituting $Re (B)$ in the initial
Eq. (2) by:
\begin{equation}
Re(B') = Re(B) - H
\end{equation}
Here $H$ stands for the value of the external magnetic field. Since the
"pseudo-magnetic" amplitude $B(E)$ is energy-dependent, we can do the
compensation by, say, putting $Re B'(E)=0$ at $E=E_p+\Gamma_p/2$. Fig. 1d
shows the dependence of relative error on $x$ with this compensation.
As expected, all the oscillations of Fig. 1b disappear and the relative
error shows a minimum at around $x\approx 2.5\lambda$.

However, the effect $T$ itself depends on the value of $Re B(E)$ - see
Eq. (8). Without the compensation $Re( B)\gg Im (B)$ (approximately by 3
orders of magnitude) and the dominant
contribution to the denominators and numerators of Eq. (8) comes from it.
If we do the above compensation, then $Im(B)\gsim Re(B)$ and
the effect in the vicinity of p-resonance
($|E-E_p| \leq \Gamma_p$) can be expressed as:
\begin{equation}
T' \approx -2 \frac{Re(D)}{Im(B)}
\end{equation}
Taking into account the energy dependence of the amplitudes, we get:
\begin{equation}
T'\approx 4\frac{\gamma^p}{\gamma^s} \cdot \frac{V_T}{\Gamma_s}
\cdot \frac{d(E-E_p)}{(E-E_p)^2+\Gamma^2/4}
\end{equation}
Therefore the effect now changes sign at around the resonance energy $E_p$
and reaches at the points $E\approx E_p\pm\Gamma_p$ its maximal value:
\begin{equation}
T' \approx \frac{\gamma^p}{\gamma^s}\cdot \frac{V_T}{\Gamma} \cdot
\frac{d}{\Gamma}
\end{equation}
Comparing this result with Eqs. (11), (12), we see that the compensating
magnetic field, besides removing the oscillations of the relative error,
also produced a important increase of the value of $T$ itself, giving an extra
resonance enhancement factor $ d/\Gamma\sim 10^3$. It also radically
changed the energy-dependence of the effect. By comparing Eqs. (8) and (16)
we see that the relative error in the presence of compensation decreases
by the same 3 orders of magnitude.

These conclusions are confirmed by the results of calculating the effect
under conditions of complete compensation $Re(B'(E_p+\Gamma_p/2))=0$
- see Fig. 1c.

Thus we see, that our initial idea [4] of compensation the pseudomagetism
turns out still to be quite productive. The only remaining point is to
estimate the practically necessary accuracy of this compensation. Following
[4], we still think that the practical way of controlling this accuracy
is by measuring the neutron spin rotation angle
$\phi=2Re(B)/Im(f)\cdot x/\lambda$ around $\vec{I}$ after its transmission
through the target sample. Fig. 2a shows the dependence of the
effect $T(E=E_p+\Gamma_p/2)$ on the spin rotation angle (which serves as a
measure of the applied compensating magnetic field).

Fig. 2b shows the same
dependence for the relative error. We see that both the effect and its
relative error are optimal for practically complete compensation
($\phi\approx 0$). The slight shift of optima to small positive $\phi$ is
caused by the interference of contributions to the effect from the
pseudo-magnetic rotation and rotation caused by the T-noninvariant field $D$.
However, the relative error changes only by a factor of 2 - 3  when
the rotation angle varies from $0^0$ to $200^0$. Thus the limitations on
the accuracy of compensation are quite moderate from this point.

A more essential limitation might come from the fact that the energy
dependence of the effect (and, to somewhat less extend, its maximal value)
changes rapidly with increasing $\phi$. In order to see this, one might
compare the curves in Fig. 1c (corresponding to $\phi=0^0$) and Fig. 1a,
calculated without compensation.

Therefore we decided to formulate the problem
of the compensation accuracy in a slightly different way: We assume that
a reasonable value for the experimental energy resolution is
$\Delta E\approx 10^{-2}$ eV and consider the practically reasonable
accuracy $\Delta \phi$ of measuring $\phi$ as a free parameter. Then the
rotation angle $\phi$ (and thus the
compensating field $H$) should be chosen in such a way that energy maximum
of the effect $T(E)$ should be shifted by less than $\Delta E$ while
varying the rotation angle in the interval from $\phi -\Delta\phi$ to
$\phi +\Delta\phi$. On performing a good deal of "computer experiments"
we can state, that the accuracy $\Delta\phi=5^0$ is quite sufficient
from this point of view.

Thus we see, that the limitations on the accuracy of measuring the rotation
angle in order to check the compensation of pseudo-magnetic rotation are
quite tolerable.\\

\noindent {\bf 3. Analysis of $\tilde{X}$.}
Consider now the quantity $\tilde{X}$. As already mentioned, it differs
from $T$ only by the normalization factor. Therefore it is also enhanced
in the vicinity of the p-wave resonance. However the new normalization
makes the effect itself (and not only its relative error) dependent both
on the angle $\theta$ and on the target thickness $x$. Moreover, the rapid
energy oscillations are superimposed on the resonance behaviour of the
effect. The character of these oscillations depend on the target thickness
$x$ in a very complicated way.  For the sake of illustration we show in
Fig. 3 the energy dependence of $\tilde{X}$, calculated for
$x \approx 2.5\lambda$.

All this considerably complicates the analysis. It is difficult even
to find a reasonable analytical approximation for $\tilde{X}$.
In the case of thick target (for $La$ resonance this means
$x\gsim 15\lambda$) one can write:
\begin{equation}
\tilde{X} \approx - \frac{\sin^2(\theta) Im(DB^*)}
{\sin^2(\theta) Im(DB^*) + 2Re\left( q (C^* + \cos(\theta)B^*)\right)}
\end{equation}
Eq. (22) shows that in the thick target limit the rapid oscillations of the
effect disappear. This makes the analysis of its energy and $\theta$
dependence much easier. Consider now the $\theta$ dependence of the
numerator and the denominator in $\tilde{X}$ separately.

The whole interval of $\theta$ values can be separated into two regions.
In the first region one can neglect all the contributions to the
denominator besides $\ 2 \cos(\theta)Re(q B^*)$. In this region
$q \approx \frac12 B$ and 
\begin{equation}
\tilde{X} \approx -\frac{\sin^2(\theta)}{\cos(\theta)}
\frac{Im(DB^*)}{|B|^2}.
\end{equation}
One can see that in this region $\tilde{X} \sim -T$.

Consider now the relative error of $\tilde{X}$ in this range of $\theta$.
In analogy to the above $T$
case, the main contribution to this error comes from the numerator. Therefore
\begin{equation}
\frac{\sigma_{\tilde{X}}}{\tilde{X}} \approx
\frac{\sigma_T}{T},
\end{equation}
This conclusion turns out to be valid even without the thick target
approximation. Therefore the relative error of $\tilde{X}$ strongly
oscillates with the variation of the target thickness.
The necessity of
the compensating external magnetic field is again obvious. Introducing this
compensation, we again observe that the oscillations of
$\sigma_{\tilde{X}}/\tilde{X} (x/\lambda)$ disappear, and it is possible to
find the optimal target thickness (which is obviously $x \approx 2.5 \lambda$).

In the first region of angles the value of the effect increases when
$\theta$ approaches the critical point where the denominator of the effect
equals zero. The relative error remains more or less constant. This does
not mean, however, that it is better to make measurements closer to
this critical point, because the absolute value of the error also
increases. Therefore the accuracy of experimental observations remains
practically the same.

The second region of angles is characterized by the inequality
$\cos(\theta)Re(q B^*) \ll Re(qC^*)$ and is located in the vicinity of
$\theta = \pi/2$. The width of this region depends on the incident
neutron energy and on the magnitude of the compensating magnetic field.
Without the compensation this width is $10^{-4} \div 10^{-7}$ rad.
In case of full compensation it increases to a few degrees.
It is important to note that this region contains the value of $\theta$
which turns the denominator into zero, while the effect formally increases
to infinity.
In this second region the denominator's contribution to the relative error
dominates. Therefore the relative error continuously increases and
becomes infinite in point where the denominator equals to zero. This
means that one should not come too close to the values  $\theta
= \frac{\pi}{2}$ because of the finite angular divergence of any
experimental beam.  It is obviously practically impossible to estimate
the actual accuracy of the measurements carried close to
$\theta = \frac{\pi}{2}$.

In order to estimate the necessary accuracy of compensation by the external
magnetic field, one should perform the same kind of analysis as in the case
of $T$. It is obvious that the results of such an analysis would be essentially
the same: the main limitation on the accuracy should again come from the rapid
change of the effect's energy dependence. As in the case of $T$,
the resulting limitations on $\Delta\phi$ accuracy are quite reasonable.\\

\noindent{\bf 4. Summary}\\
We can draw the following conclusions:

Analysis of the CP-violating effect's relative error is by no means less
essential than analyzing the effect itself. One can always normalize the
CP-noninvariant difference (3) dividing it by a very small quantity. However
such a normalization would not increase the accuracy of the measurement.

The necessity to compensate the pseudomagnetic precession is caused
essentially by the fact that without such a compensation the accuracy
of measurement varies with target thickness in a practically uncontrollable
way.

The compensation of the pseudomagnetic precession  increases by 3 orders
of magnitude not only the effect itself but also the accuracy of its
measurement. The net enhancement in the vicinity of p-wave resonance
with compensation reaches 6 orders of magnitude. The energy dependence
of the effect changes drastically in the presence of compensation.

As a practical way to control the degree of compensation we suggest,
following [3], to measure the rotation angle $\phi$ of neutron
polarization around the target polarization vector. When $\phi$
varies between $0^0$ and $200^0$ the maximal value of the effect and
its relative error varies not more than by a factor of $2\div 3$.
The most stringent restriction on the accuracy of measurement of this
angle comes from the fact that the energy dependence of the measured effect
strongly depends on the value of the compensating magnetic field. With this
restriction in mind it seems sufficient to fix $\phi$ with the accuracy about
$5^0$.

The CP-noninvariant quantity $X$ suggested for measurement in ref. [7] is
shown to differ from Stodolsky's CP-noninvariant difference (3) practically
only by the choice of normalization factor. This factor becomes zero in the
vicinity of $\theta=\pi/2$ (the beam polarization orthogonal to the target
one). Although the value of thus normalized effect tends to infinity, its
relative error also tends to infinity in this range of $\theta$ values.
In the remaining range of $\theta$, where the normalizing factor exceeds
the CP-noninvariant difference (3), the relative error depends on $\theta$
angle as $1/\sin(\theta)$.  Although the value of the effect in this range of
$\theta$ angles behaves as $\tan(\theta)$ and strongly increases approaching
the value $\theta\approx \pi/2$, the relative accuracy of its measurements
(besides the small vicinity of $\theta\approx 0$) remains practically
the same.

We acknowledge the support of the Russian Fund of Fundamental Studies
(grant No. 97-02-16803).

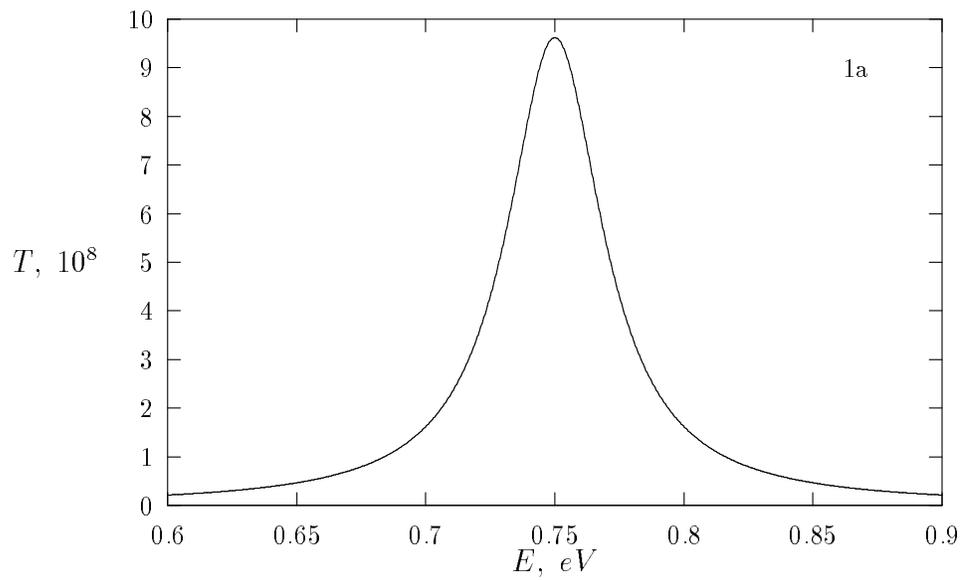
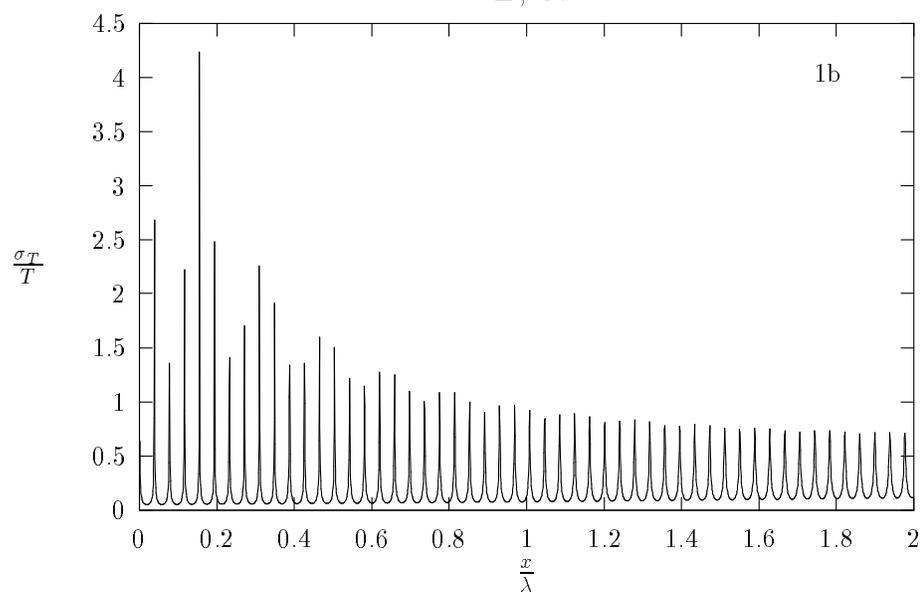
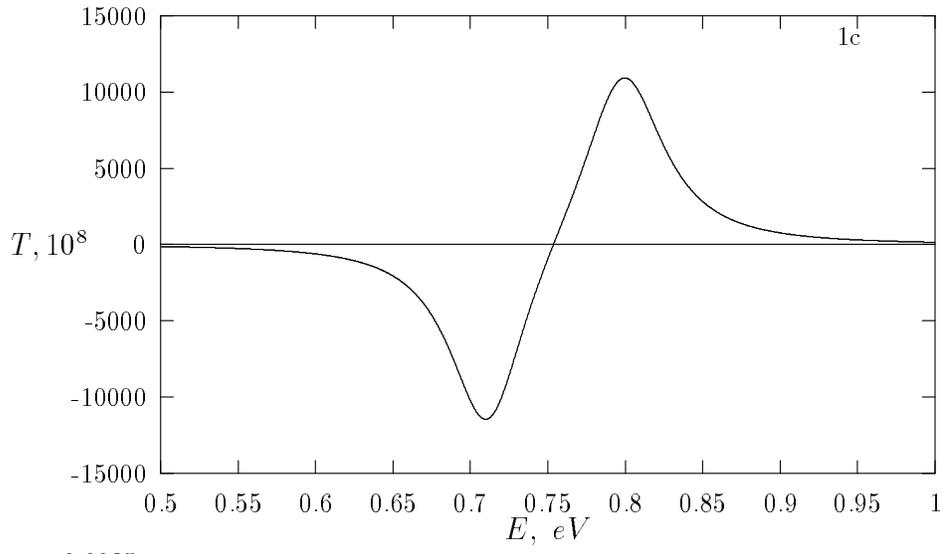
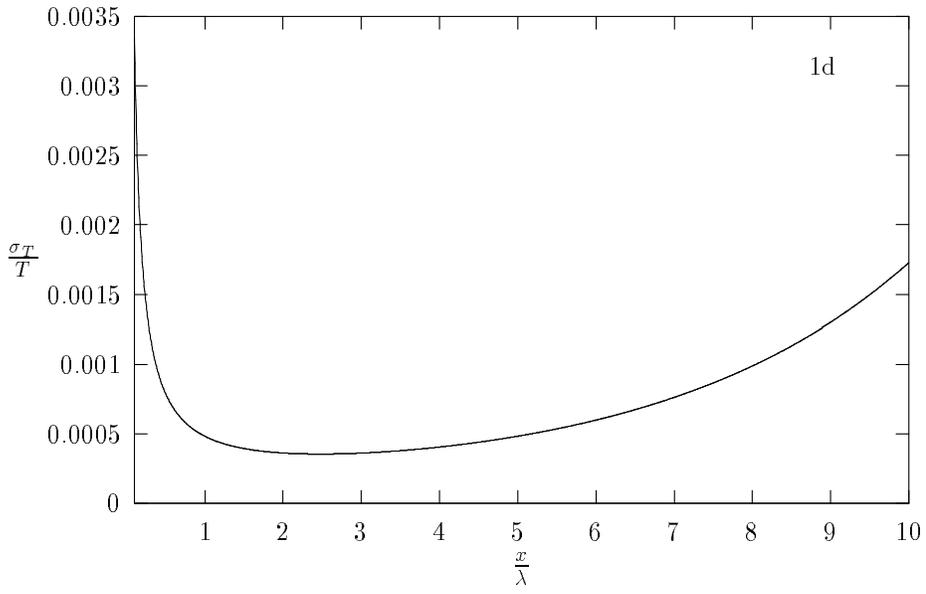
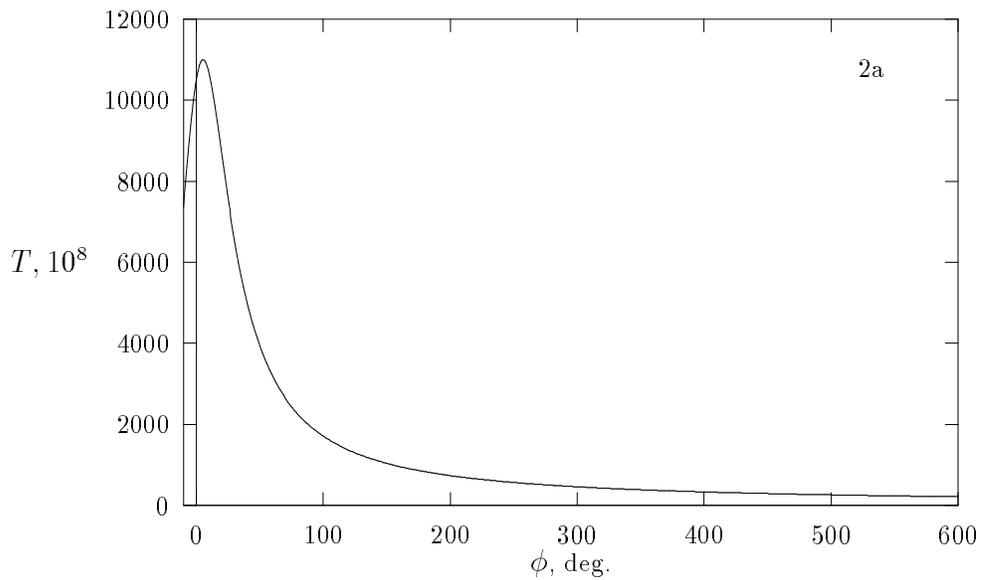
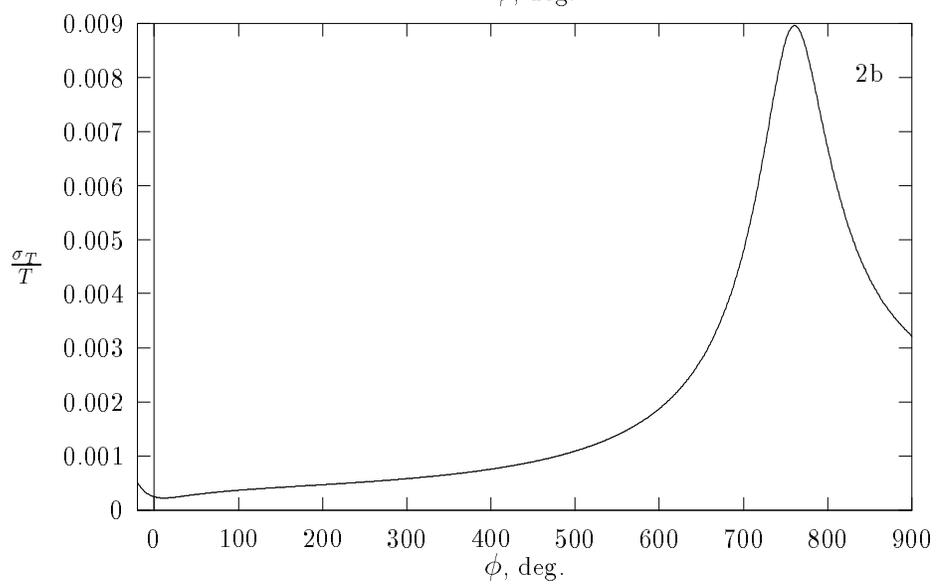
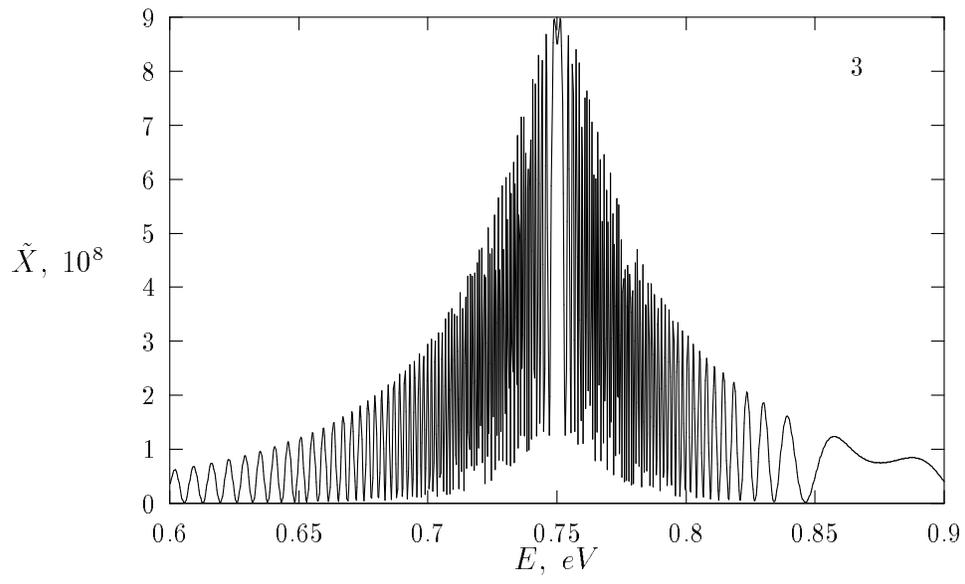
\end{document}